\documentstyle[12pt]{article}

\begin{document}
\title{
\begin{center}
\textbf{\large THE NEWTONIAN CORRESPONDENCE}
\end{center}}
\author{Amir H. Abbassi\thanks{%
E-mail: ahabbasi@net1cs.modares.ac.ir}$\;\;\;\&\;$ Amir M. Abbassi\thanks{%
Permanent add: Depatment of Physics, Faculty of Sciences, Tehran University.}$%
\;^,\;$\thanks{%
E-mail: amabasi@khayam.ut.ac.ir} \\
\textit{\small Department of Physics, School of Sciences, }\\
\textit{\small Tarbiat Modarres University,P.O.Box 14155-4838,}\\
\textit{\small Tehran, Iran.}}


\date{Dec. 2001}
\maketitle 
\bigskip

It is shown that the field equations of general theory of relativity in the Einstein
tensor form and the unimodular theory of gravity do not fulfill the
correspondence principle commitment completely. The  consistent
formalisms are briefly discussed.\\ 

\noindent Key words:  correspondence principle,  general relativity, unimodular gravity.                 \\

\newpage
\bigskip 
\noindent\textbf{1. INTRODUCTION}\\

\smallskip
General relativity (GR) is established upon some physical principles. The principle of covariance (PC) and the
correspondence principle (CP) are two of them which are considered in this letter. According to PC the field equations should have tensorial form,
and GR must agree with the Newtonian gravitational theory in
the limit of weak gravitational fields and low velocities by CP. 
Different aspects of the Newtonian
limit may be classified as follows:

\begin{itemize}
\item[a]  - The equation of geodesic deviation.

\item[b]  - The geodesic equation.

\item[c]  - The weak field limit of GR should give the same equations of
motions as Newtonian gravity.
\end{itemize} 

In GR we are dealing with second rank tensorial field equations,
generally a set of ten relations, while in the Newtonian gravity we have only one Poisson
equation and it seems there is no correspondence for nine 
of the rest. Does this mean that the PC breaks in taking the Newtonian
limit? The answer is negative. In the Newtonian limit the
Lorentz transformations reduce to Galileo transformations,
so that $t$ appears as a scalar. By Newtonian correspondence 
we must consider weak fields and low velocities. It turns out that
in the spatial components of the field equations the first non-zero
term has an order of approximation higher than the corresponding one in the $tt$-component. Since in finding the Newtonian limit 
we merely keep the first order terms in the $tt$-component, this
leads to $0=0$ for other components.
For more clarification we may work in a system of units that
$c\neq 1$. This explicitly shows , when the velocity 
of light tends to infinity, how some components of the
field equation disappear.
From this point of view we may say that PC is not violated but the other components have 
no physical information. So we may restate the item (c) as follows: 

\begin{itemize}
\item[\'c]  - The (00)-component of the field equation must reduce to the
Poisson equation for a weak stationary field produced by nonrelativistic
matter[1].
\end{itemize}
We are going to show that, from (\'c) point of view, the GR field equations in the form of Einstein tensor and the field equations of the unimodular gravity
do not satisfy CP. The consistent form and 
its consequences are discussed.  \\

\bigskip
\noindent\textbf{2. EINSTEIN TENSOR FORM}\\

\smallskip
 We restrict our discussion to the Schwarzschild space which is the solution of
the field equations for spherically symmetric vacuum space around a point
mass M. In the literature we have the Einstein field equations in the form
Einstein tensor proportional to energy-momentum tensor i.e.  :

\begin{equation}
R_{\mu\nu}-\frac{1}{2}g_{\mu\nu}R=-\frac{8{\pi}G}{c^4}T_{\mu\nu}   \label{1}
\end{equation}
satisfying CP so that the (00)-component of the field equation reduces to the
Poisson equation in the weak field limit. It will be shown that  for the Schwarzschild metric this is not so.

The Weinberg's argument to reach this result is based on the
fact that in a nonrelativistic system $T_{ij}\ll T_{00}$ , then $%
|G_{ij}|\ll |G_{00}|$ and $R_{ij}\approx \frac{1}{2}g_{ij}R$. Furthermore $%
g_{\alpha\beta}\approx\eta_{\alpha\beta}$ and the curvature scalar is given
by
\begin{equation}
R\approx R_{kk}-R_{00}\approx 2R_{00}  \label{2}
\end{equation}
So concludes that $G_{00}\propto R_{00}$ and $R_{00}\propto \nabla^2g_{00}$  [2] .
The weak point in this argument is that by making use of
$G_{ii}=0$ in calculating $G_{tt}$ actually different
components of the field equations are combined. In other 
words the Poisson equation is constructed by a proper
mixing of all the available equations. This is in contrast 
with the original claim that the $(tt)$-component of the 
field equation in the weak field limit gives the Poisson equation.

The Einstein tensor in the weak field limit may yield to$\;$ $\Box \psi_{\mu\nu}
$ , $\;$ where$\;$ $\psi_{\mu\nu}\equiv h_{\mu\nu}-\frac12 \eta_{\mu\nu}h\;\;,\;h_{\mu\nu}= g_{\mu\nu}-{\eta}_{\mu\nu}$ in Minkowski coordinate , provided that
the Einstein condition is satisfied as follows [3] 
\begin{equation}
h_{\nu,\mu}^\mu-\frac{1}{2}h_{,\nu}=0 \;\;\;,\;\;\;h=\eta^{\mu\nu}h_{\mu\nu}  \label{3}
\end{equation}
If this condition holds ,  in the stationary case $\Box$ reduces to $\nabla^2$
and the Poisson equation is obtained automatically. Let us see what happens 
in the weak field limit of Schwarzschild metric. We have

\begin{eqnarray}
h_{tt}=\frac{2\phi}{c^2}\;\;\; , h_{x_ix_i}=\frac{2\phi {x_i} ^2}{r^2 c^2}\;\;\; ,
h_{x_ix_j}=\frac{2\phi x_i x_j}{r^2 c^2}\;\;\;\;,\;\; i,j=1,2,3\nonumber\\
-\phi=\frac{GM}{r}\ll c^2 .\qquad\qquad\;\;\;\;\;\;\;\;\;\;\;\;\;\;\;\;\label{4}
\end{eqnarray}
Using (4) we get $h=0$ and these do not satisfy (3), i.e. Einstein condition 
does not hold in this case. It means that for Schwarzschild space we do not end 
to the Poisson equation in the weak field limit.\\
Since curvature tensor and its contractions are invariant quantities under a
gauge transformation of $ h_{\mu\nu}$ as follows
\begin{eqnarray}
x^\mu\rightarrow {x^{^\prime}}^\mu=x^\mu +\epsilon \xi^\mu  \qquad\nonumber\\
h_{\mu\nu}\rightarrow {h^{^\prime}}_{\mu\nu}=h_{\mu\nu}-
2\xi_{(\mu,\nu)}   \label{5}
\end{eqnarray}
it is possible to find a gauge in which Einstein condition holds. This gauge may be
obtained from 
\begin{equation}
\Box \xi_\nu={\psi^\mu}_{\nu,\mu}\;\;\;\;,\;\;\; {\psi^\mu}_{\nu,\mu}
={h^\mu}_{\nu,\mu}-\frac12h_{,\nu}  
\end{equation}  \label{6}
We may conclude that the weak field limit of GR and Newtonian field equation
are not in the same gauge.

In what follows we will see that this approximation although may lead
to a correct prediction of reciprocal of distance for Newtonian potential but indeed
 does not reduce to the Poisson equation as is required.

The line element of a spherically symmetric vacuum space is
\begin{equation}
ds^2=B(r)c^2 dt^2-A(r)dr^2-r^2(d\theta ^2+\sin^2\theta d\varphi^2)  \label{7}
\end{equation}
where $B(r)=A^{-1}(r)=1+\frac{2\phi}{c^2}$. Using (7) the
nonvanishing components of Einstein tensor are:
\begin{equation}
G_{rr}=-\frac{B^{^{\prime}}}{rB}+\frac{A-1}{r^2}  \label{8}
\end{equation}
\begin{equation}
G_{\theta\theta}=-\frac{r^2B^{^{\prime\prime}}}{2AB}+\frac{r^2B^{^{\prime}}}{%
AB}(\frac{A^{^{\prime}}}A+\frac{B^{^{\prime}}}B) -\frac r{2A}(-\frac{%
A^{^{\prime}}}A+\frac{B^{^{\prime}}}B)  \label{9}
\end{equation}
\begin{equation}
G_{\varphi\varphi}=\sin^2\theta\;G_{\theta\theta}  \label{10}
\end{equation}
\begin{equation}
G_{tt}=c^2[-\frac{BA^{^{\prime}}}{rA^2}+\frac B{r^2}(-1+\frac 1A)]  \label{11}
\end{equation}
prime stands for differentiation with respect to $r$.

In contrast to what is expected, $G_{tt}$ for the Schwarzschild metric merely
contains the first order differentiation with respect to $r$ and in no way
can yield to the Poisson equation in weak field limit. Therefore there is
an obvious discripancy between the obtained result and the Newtonian equation.
Although (11) in the limit of weak fields gives 
\begin{equation}
G_{tt}\cong2(\frac{\phi^{^{\prime}}}r+\frac \phi{r^2})  \label{12}
\end{equation}
which has the same solution of reciprocal of $r$ as the Poisson equation possess
for a particle with mass M. This can be considered as a gauge violation of CP
which may be forbidden too. \\

\bigskip
\noindent\textbf{3. UNIMODULAR GRAVITY} \\

\smallskip
In a more plausible consideration of cosmological constant as an integration
constant the unimodular gravity is actually very well motivated. If the
determinant of $g$ is not dynamical then the action only has to be
stationary with respect to variations in the metric for which $g^{\mu\nu
}\delta g_{\mu\nu}=0$ , yielding the field equations  [4,5,6] 
\begin{equation}
R^{\mu\nu}-\frac 14 g^{\mu\nu}R=-\frac{8\pi G}{c^4}(T^{\mu\nu}-\frac 14 g^{\mu
\nu}T_\alpha ^\alpha)  \label{13}
\end{equation}
with $T^{\mu\nu}$ as conserved stress tensor of matter. The combination of
this with Bianchi identities for the covariant derivative of the Einstein tensor
gives a nontrivial consisting condition 
\begin{equation}
\frac14 \partial_\mu R=\frac{8\pi G}{c^4} (\frac14 \partial_\mu T_\lambda^\lambda)
\label{14}
\end{equation}
Denoting the constant of integration by $-4\Lambda$ the Einstein field
equations is recovered.

We also see that this form of field equations i.e. (13),  regretfully does not
satisfy the CP from (\'c) point of view. For spherically symmetric vacuum space (7) the components of (13) are : 
\begin{equation}
R_{rr}-\frac14 g_{rr}R=\frac{B^{^{\prime\prime}}}{4B}-\frac{B^{^{\prime }}}{%
8B}(\frac{A^{^{\prime}}}A+\frac{B^{^{\prime}}}B)+\frac{A-1}{2r^2} -\frac{%
A^{^{\prime}}}{rA}  \label{15}
\end{equation}
\begin{equation}
R_{\theta\theta}-\frac14 g_{\theta\theta}R=-\frac{r^2B^{^{\prime\prime }}}{%
4AB}+\frac{r^2B^{^{\prime}}}{8AB}(\frac{A^{^{\prime}}}A+ \frac{B^{^{\prime}}}%
B)+\frac 12(\frac 1A-1)  \label{16}
\end{equation}
\begin{equation}
R_{\varphi\varphi}-\frac 14 g_{\varphi\varphi}R=\sin ^2\theta\;(R_{
\theta\theta}-\frac 14 g_{\theta\theta}R)  \label{17}
\end{equation}
\begin{eqnarray}
R_{tt}-\frac 14 g_{tt}R=c^2[-\frac{B^{^{\prime\prime}}}{4A}+\frac{B^ {^{\prime}}%
}{8A}(\frac {A^{^{\prime}}}A+\frac{B^{^{\prime}}}B)- \frac B{2rA}(\frac{%
A^{^{\prime}}}A+\frac{B^{^{\prime}}}B)  \nonumber \\
-\frac B{2r^2}(1-\frac 1A)]  \label{18}
\end{eqnarray}
In the weak field limit for the (18) we get 
\begin{equation}
R_{tt}-\frac 14 g_{tt}R=-\frac{\phi^{^{\prime\prime}}}2+ \frac \phi{r^2}
\label{19}
\end{equation}
Again for (19) we have reciprocal of $r$ as its solution but it is not the
Poisson equation as is expected from CP.\\ 

\bigskip
\noindent\textbf{4. CONSISTENT FORM} \\

\smallskip
The ordinary field equations in the following form 
\begin{equation}
R_{\mu\nu}=-\frac{8\pi G}{c^4}(T_{\mu\nu}-\frac 12 g_{\mu\nu}T)  \label{20}
\end{equation}
fulfill the CP requirement , that is the $(00)$- component of (20) for weak
field limit of Schwarzschild metric reduces to 
\begin{equation}
-\phi ^{^{\prime\prime}}-\frac {2\phi ^{^{\prime}}}r=-4\pi GM \delta(\vec{r})
\label{21}
\end{equation}
which in a compact form is exactly the Poisson equation 
\begin{equation}
\nabla ^2\phi=4\pi GM\delta(\vec{r})  \label{22}
\end{equation}

For a perfect fluid the (00)-component of the RHS 
of (20) in the weak field limit reduces to
\begin{equation}
4\pi G(\rho +3p/c^2)  \label{23}
\end{equation}
which is equal to $8\pi G \rho_t$ where 
$\rho_t$ is the timelike convergence density [7].
In the limit of slow motion , $\rho\gg p/c^2$,  and
$p/c^2$ 
can be ignored so that $\rho_t =\rho /2$, and Eq.(22)
gives
\begin{equation}
\nabla^2 \phi = 4 \pi G \rho  \label{24}
\end{equation}

The reason why this discripancy has not been recognized is that in finding
the Schwarzschild metric we usually solve $R_{\mu\nu}=0$ as field equation.
We may conclude that the form of the Einstein field equations with cosmological
constant consistent with CP is 
\begin{equation}
R_{\mu\nu}=-\frac{8\pi G}{c^4}(T_{\mu\nu}-\frac 12 g_{\mu\nu}T)-\Lambda g_{\mu\nu} 
\label{25}
\end{equation}

This field equation may be derived from standard actions by considering the
density metric of weight $+1$ instead of the metric as dynamical variables
which is defined as  [8]: 
\begin{equation}
\tilde{g}_{\mu\nu}=\sqrt{-g}\;g_{\mu\nu}  \label{26}
\end{equation}
and we get 
\begin{equation}
\delta I=\int d^4x\{\frac {c^4}{16\pi G}(R^{\mu\nu}+\Lambda g^{\mu\nu })+\frac
12(T^{\mu\nu}-\frac 12 g^{\mu\nu}T)\}\delta \tilde{g}_{\mu\nu}  \label{27}
\end{equation}
From (26) we have 
\begin{equation}
\delta \tilde{g}_{\mu\nu}=\sqrt{-g}\delta g_{\mu\nu}-\frac 12\sqrt{-g}
\;g_{\mu\nu}g^{\alpha\beta}\delta g_{\alpha\beta}  \label{28}
\end{equation}
Inserting (28) in (27) gives the ordinary variation of standard action with
respect to the variation of the metric. 
\begin{equation}
\delta I=\int d^4x\{\frac {c^4}{16\pi G}(R^{\mu\nu}-\frac 12 g^{\mu\nu
}R+\Lambda g^{\mu\nu})+\frac 12T^{\mu\nu}\}\sqrt{-g}\delta g_{ \mu\nu}
\label{29}
\end{equation}
This procedure may be carried out in an elegant way by applying the Palatini
approach based on the idea of treating the metric (the density metric) and
the connection separately as dynamical variables which the variation with
respect to the connection reveals that the connection is necessarily the
metric connection.\\
It is evident from (29) that the common field equations (1) are obtained 
under the variations of $\delta g_{\mu\nu}$ with the condition that$\mid g
\mid\not=0$. While the consistent form (20) are resulted from (27) under the 
variations of $\delta\tilde g_{\mu\nu}$ without any condition.\\

\bigskip
\noindent\textbf{5. REMARKS}  \\

\smallskip
Let us  summarize the significant results.
\begin{itemize}
\item[1]- It is shown that how (\'c) statement may be explicitly obtained
from (c) statement in the mentioned CP classification without violating PC.
\item[2]- Einstein field equations in the common form (1) of Einstein
tensor proportional to the energy-momentum tensor do not fulfill the 
CP from (\'c) point of view.
 \item[3]- The unimodular gravity field equations (13) do not satisfy the (\'c)  statement.
\item[4]- The alternative field equations (20) which are mathematically
equivalent to the Einstein common field equations (1) satisfy the CP
commitments completely. This means that indeed these two forms are
not physically equivalent. In Cartesian spatial coordinates 
the Poisson equation may be obtained from all the components
of this form of field equations.
\item[5]- The failure of unimodular model in this study ceases the
interpretation of the cosmological constant as an integration constant,
i.e. it is a universal constant of nature.
\item[6]- Derivation of Eq.(1) from Lagrangian formalism (29) requires
the constraint $\mid g\mid\not=0$. Thus the resulted field equations 
are restricted and are not necessarily defined for the whole space.
\item[7]- By taking the density metric tensors (26) as  dynamical variables
the obtained field equations from Lagrangian formalism (27) are free from
any constraint and holds everywhere. 
\end{itemize}

Accordingly, we should accept to carry out recasting of 
the GR field equations.\\

\end{document}